\documentclass[showpacs,showkeys,prl,twocolumn,aps,preprintnumbers,amsmath,amssymb]{revtex4}

\usepackage[breaklinks=true,colorlinks,citecolor=blue,linkcolor=blue,urlcolor=blue]{hyperref}
\usepackage{graphicx,dcolumn,bm,multirow,hyperref,mathtools,braket,color,url,epstopdf,amssymb,amsmath}

\def\Ef{$E_f$}

\begin{document}


\title{Fermi surface studies of a non-trivial topological compound YSi}

\author{Vikas Saini}

\author{Souvik Sasmal}

\author{Ruta Kulkarni}

\author{Bahadur Singh}

\author{A. Thamizhavel}
\email{thamizh@tifr.res.in}
\affiliation{Department of Condensed Matter Physics and Materials Science, Tata Institute of Fundamental Research, Homi Bhabha Road, Colaba, Mumbai 400005, India.}

\date{\today}

\begin{abstract}
	
The Fermi surface properties of a nontrivial system YSi is investigated by de Haas-van Alphen (dHvA) oscillation measurements combined with the first-principle calculations. Three main frequencies ($\alpha$, $\beta$, $\gamma$) are probed up to $14$~T magnetic field in dHvA oscillations. The $\alpha$-branch corresponding to $21$~T frequency possesses non-trivial topological character with $\pi$ Berry phase and a linear dispersion along $\Gamma$ to $Z$ direction with a small effective mass of $0.069~m_e$ with  second-lowest Landau-level up to $14$~T.  For $B~\parallel$~[010] direction, the 295~T frequency exhibits non-trivial $2D$ character with $1.24\pi$ Berry phase and a high Fermi velocity of $6.7 \times 10^5$~ms$^{-1}$.  The band structure calculations reveal multiple nodal crossings in the vicinity of Fermi energy $E_f$ without spin-orbit coupling (SOC).  Inclusion of SOC opens a small gap in the nodal crossings and results in nonsymmorphic symmetry enforced Dirac points at some high symmetry points, suggesting YSi to be a symmetry enforced topological metal.  

\end{abstract}

\maketitle

\section{Introduction}

Topological semimetals have gained significant attention both on theoretical and experimental fronts in recent years owing to their exotic properties. They exhibit symmetry protected bulk and surface states which results novel phenomena and transport properties with tremendous potential for applications in quantum technologies and energy sciences ~\cite{Bansil16,Hasan10,Qi11,  vergniory2019complete, hu2019transport, fu2007topological}. Topological semimetals can be classified based on the dimensionality and degeneracy of the crossing points. In particular, topological Dirac semimetal has zero-dimensional four-fold degenerate crossing points in the vicinity of Fermi energy (\Ef).  The low energy excitations of the so-called Dirac particles follow linear energy dispersion around the Dirac point and are the reasons for the observed exotic phenomenon in the electrical transport measurements. The four-fold degeneracy of the Dirac point is ensured by the presence of both the  time-reversal $\mathcal{T}$ and inversion $\mathcal{I}$ symmetries along with specific crystalline symmetries. This four-fold degeneracy is lifted by breaking either inversion or time-reversal symmetries, thus  Dirac semimetals (DSM)  transitions to  Weyl semimetals (WSM) with two-fold Weyl nodal crossings~\cite{singh12,wang2013three, wang2012dirac, lv2015observation}. Cd$_3$As$_2$ and Na$_3$Bi are known prototype topological Dirac semimetals which was predicted theoretically and realized in experiments. They exhibit ultra-high mobility, extremely high magnetoresistance and show exotic phenomena such as chiral anomaly, quantum hall effect, among others owing to symmetry protected Dirac band crossings. The extensive studies of Shubnikov de-Haas(SdH) and de Haas-van Alphen(dHvA) oscillations confirmed topological nontrivial nature of these compounds from the Berry phase analysis and other unusual properties~\cite{wang2013three, wang2012dirac, xu2013, zhang2019quantum, zhang2017room, kushwaha2015bulk, chiu2015classification, liang2015ultrahigh}. 

In view of finding new topological semimetals, we investigate Fermi surface and topological properties of binary compound YSi by de Haas-van Alphen (dHvA) oscillations and first-principles calculations. We synthesize high-quality single crystals of YSi using the Czochralski method. From dHvA quantum oscillations studies, YSi is found to have nontrivial Fermi pockets. The first-principle calculations show that it has a rich nodal structure at the Fermi level with symmetry protected band crossings. Three different branches ($\alpha, \beta$, and $\gamma$) are probed up to a magnetic field of $14$~T. The nontrivial $\alpha$ pocket is observed along the three directions with a subtle variation in the frequency implying a non-uniform cross-section of the Fermi surface suggesting an anisotropic nature of the $\alpha$ pocket. On the other hand, the other two pockets ($\beta$ and $\gamma$) are observed only along $B~\parallel$~[010] direction suggesting a $2D$ nature of the Fermi surface, where $\beta$ branch has the nontrivial character of topology. The detailed band structure analysis shows that both the valence and conduction bands participate in the formation of the Fermi surface with the hole and electron pockets. The type-I nodal anti-band crossing along the $\Gamma-Z$ direction in the vicinity of Fermi energy leads to the nontrivial $\alpha$ pocket.  

\section{Methods}

From the binary phase diagram of Y and Si, it is obvious that YSi melts congruently at 1845~$^{\circ}$C~\cite{button1990preparation} and hence can be grown directly from its melt. A tetra-arc furnace has been used to grow the single crystal of YSi by the Czochralski method. High purity starting materials of Y (3N pure, Alfa-Aesar) and Si (5N pure, Alfa-Aesar) were taken in the ratio of $1:1.05$ and repeatedly melted to make a homogenous polycrystalline ingot of about 8 to 10 g. A seed crystal was cut from this polycrystalline ingot to grow the single crystal in a vacuum chamber filled with Ar gas. The polycrystalline ingot was melted and the seed crystal was slowly inserted into the melt and pulled very rapidly to start with. Once the steady-state condition is achieved the pulling rate was maintained at $10$~mm/h. As grown pulled ingot had a diameter $\approx 4$~mm and length $\approx 70$~mm and is shown in Fig.~\ref{Fig1}(a). The composition analysis was performed using energy dispersive analysis by x-rays (EDAX). A small portion of the crystal was crushed to fine powders and subjected to room temperature powder x-ray diffraction (XRD) measurement in PANalytical x-ray diffractometer equipped with a monochromatic Cu-$K_{\rm \alpha}$ source with the wavelength $\lambda$~=~1.5406~\AA. To confirm the single-crystalline nature of the sample and to orient and cut the crystal along the three principal crystallographic directions we have performed Laue diffraction in the back reflection geometry. The oriented crystal was cut into a rectangular bar shape using a spark erosion cutting machine. Magnetic measurements were performed in a vibrating sample magnetometer (VSM) (PPMS, Quantum Design, USA), down to $2$~K and in a magnetic field of $14$~T.

Band structure calculations were carried out with the projector augmented wave (PAW) method\cite{paw} within the density functional theory (DFT)\cite{kohan_dft} framework as implemented in the Vienna ab initio simulation package (VASP)\cite{kresse99,kresse96}. The exchange-correlation effects were considered with the generalized gradient approximation (GGA) with the Perdew–Burke–Ernzerhof parameterization\cite{pbe}. The spin-orbit coupling (SOC) was included self-consistently. An energy cut-off of 310 eV was used for the plane-wave basis set and a $11 \times 11\times13$ $\Gamma$-centered {\it k} mesh was used for the bulk Brillouin zone sampling. The Xcrysden program was used to visualize the Fermi surface\cite{xcrysden}. The robustness of results is further verified by calculating electronic properties using the WIEN2K code which considers a full-potential linearized augmented plane-wave formalism~\cite{blaha2001wien2k}. The quantum oscillations calculation were performed using the SKEAF code\cite{skeaf}. 

\section{Results and Discussion}

\subsection{Crystal Structure}

\begin{figure}[!]
\centering
\includegraphics[width=0.48\textwidth]{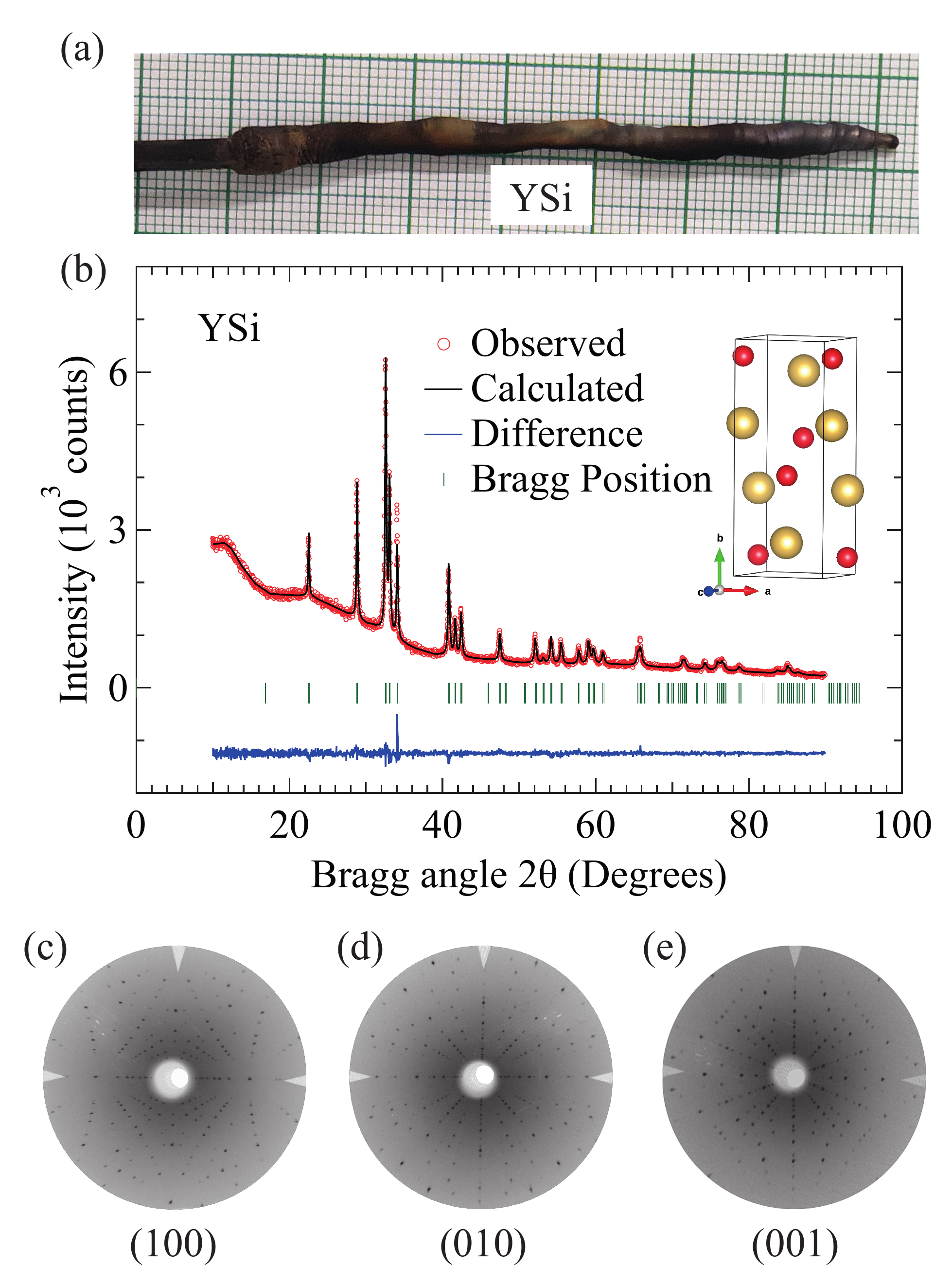}
\caption{(a) As grown single crystalline pulled ingot. (b)  Room temperature powder x-ray diffraction and structural refinement by Rietveld method. The inset shows crystal structure of YSi with $b$ as the longest axis. The large (yellow) and small (red) balls identify Y and Si atoms, respectively.  (c), (d), and (e) show the observed Laue diffraction pattern along the three principal crystallographic planes (100), (010), and (001), respectively.  }
\label{Fig1}
\end{figure}

We start discussing the crystal structure of our grown single crystals of YSi. The phase purity of the grown crystal was analysed by powder XRD measurement. It has been reported that YSi crystallizes in the centrosymmetric orthorhombic crystal structure with space group $Cmcm$ (\#63)~\cite{parthe1959crystal}. The unit cell structure of YSi is shown in the inset of Fig.~\ref{Fig1}(b).   The powder XRD pattern is shown in Fig.~\ref{Fig1}(b) and it is evident that there are no discernible impurity peaks in the entire $2\theta$ rage from 10 to 90~$^{\circ}$. From the Rietveld analysis using FULLPROF software~\cite{rodriguez1993recent} we confirmed that this compound crystallizes in the $Cmcm$ space group. The obtained lattice parameter from the Rietveld analysis are $a = 4.260$~\AA, $b = 10.530$~\AA~ and $c = 3.830$~\AA, which is in close agreement with the available data~\cite{parthe1959crystal}. The Y and Si atoms occupy $4c$ Wyckoff position with coordinates (0, 0.3598, 0.25) and (0, 0.0764, 0.25), respectively. The Laue diffraction pattern attests good quality of the grown crystal as shown in Fig.~\ref{Fig1}(c), (d), and (e) respectively for (100), (010), and (001) planes.  

\begin{figure}[h]
\centering
\includegraphics[width=0.48\textwidth]{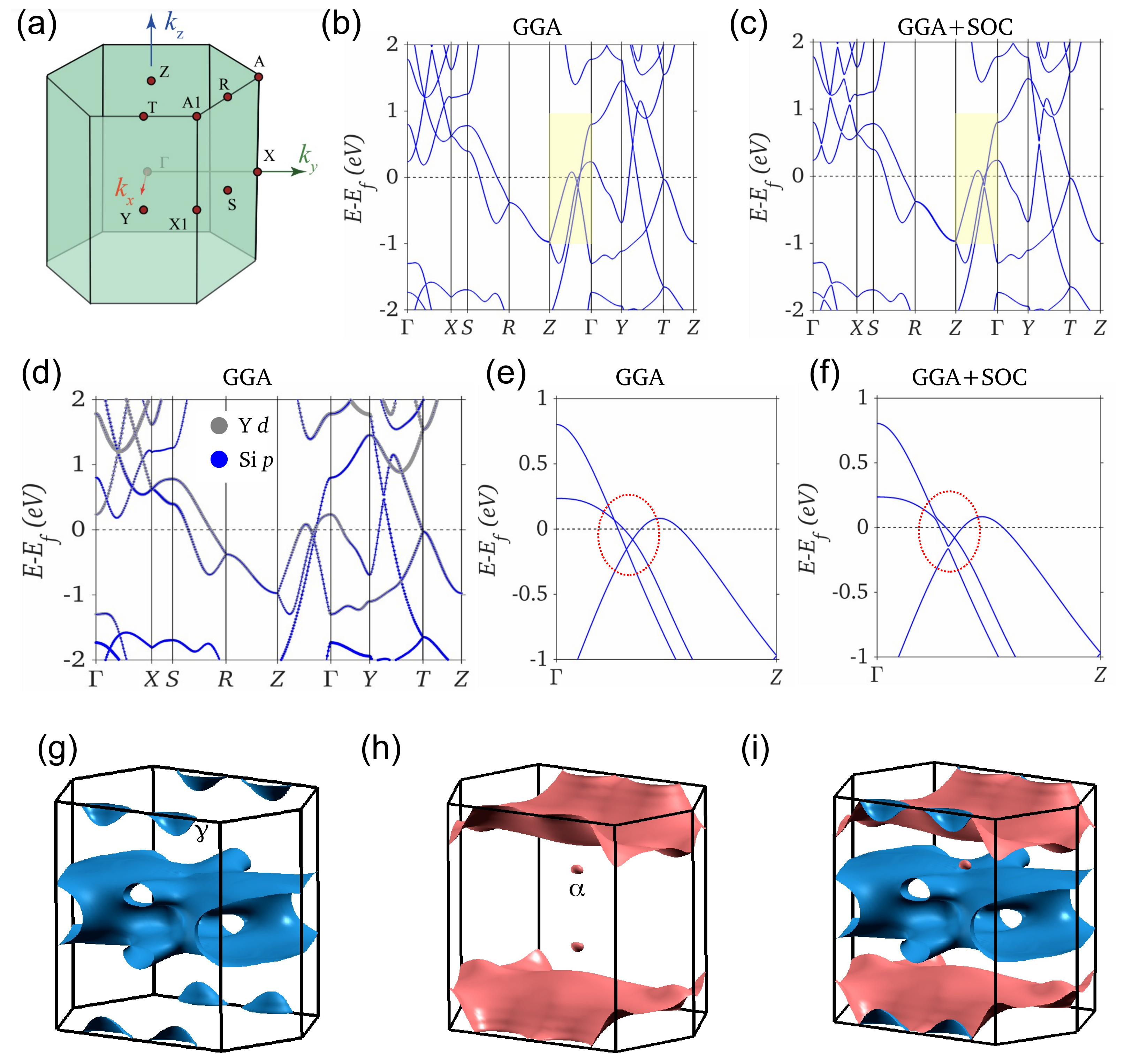}
\caption{(a) Bulk Brillouin zone of orthorhombic YSi. The high-symmetry points are marked. Bulk band structure of YSi (b) without and (c) with spin-orbit coupling (SOC). (d) Orbital resolved band structure without SOC. Y $d$ and Si $p$ states are shown with gray and blue markers, respectively. (e)-(f) Closeup of the bands along $\Gamma-Z$ direction in the area highlighted by yellow rectangles in (b)-(c). The broken circles highlight the three crossing bands. The crossing bands are gapped in the presence of SOC. (g)-(h) The calculated individual Fermi pockets and (i) the Fermi surface of YSi. }
\label{Fig2}
\end{figure}

\subsection{Electronic structure}

Figure \ref{Fig2}(a) shows the bulk Brillouin zone for the primitive crystal structure of YSi where high-symmetry points are marked explicitly. The calculated bulk band structure of YSi along various high-symmetry directions without SOC is shown in Fig. \ref{Fig2}(b). It is seen to be metallic where various bands cross the Fermi level. Importantly, many symmetry protected spinless band crossings are found along the high-symmetry lines such as $R-Z$ and $T-Z$ at the Brillouin zone boundaries as well at the generic $k-$ points. Along the $\Gamma-Z$ direction three bands cross, forming both the type-I and type-II nodal crossings as shown in Fig. \ref{Fig2}(e). The orbital resolved band structure shows that these band crossings are composed by Y $d$ and Si $p$ orbitals (Fig. \ref{Fig2}(d)). The structure with SOC is shown in Figs. \ref{Fig2}(c) and (f). The various nodal crossings at the generic $k$ points are gapped. However owing to the presence of screw rotations $\{C_{2z}|0 0 \frac{1}{2}\}$ and $\{C_{2y}| 0 0 \frac{1}{2}\}$ and glide mirror $\{M_{y}|0 0 \frac{1}{2}\}$, the band crossings at $R$, $Z$, and $T$ points remain protected, referring high symmetry point Dirac states. 

The calculated bulk Fermi surface of YSi is shown in Fig. \ref{Fig2}(i) whereas its constituents individual pockets are illustrated in Figs. \ref{Fig2}(g) and (h). Owing to the multi-band crossings at the Fermi level, the Fermi surface is composed of both the electron and hole bands. We mark a small pocket along the $\Gamma-Z$ direction as $\alpha$ which is formed by bands highlighted in the broken red circle in Fig. \ref{Fig2}(f). On the other hand, the pocket along the $T-A1$ direction is identified as $\gamma$. Additionally, giant Fermi pockets are enclosing $\Gamma$ and $Z$ points. The calculated quantum oscillations frequencies are summarized in Table~\ref{Tab1}. It is found that a shift of $\sim+11$~meV in the Fermi level is essential to reproduce the experimentally observed frequencies. The observed oscillations of the $\alpha$ pocket are observed in all three directions and it carries the lowest frequency of $21$~T for $B~\parallel~$[100]. $\beta$ pocket is observed only for B $||$ [010] and has a frequency of $61$~T. Also, the $\gamma$ pocket was observed when B $||$ [010] with a high frequency of $295$~T. These oscillations are well captured in our first-principles results  (see below for more details).

\subsection{de Haas-van Alphen quantum oscillations studies}

\begin{figure}[!]
\centering
\includegraphics[width=0.5\textwidth]{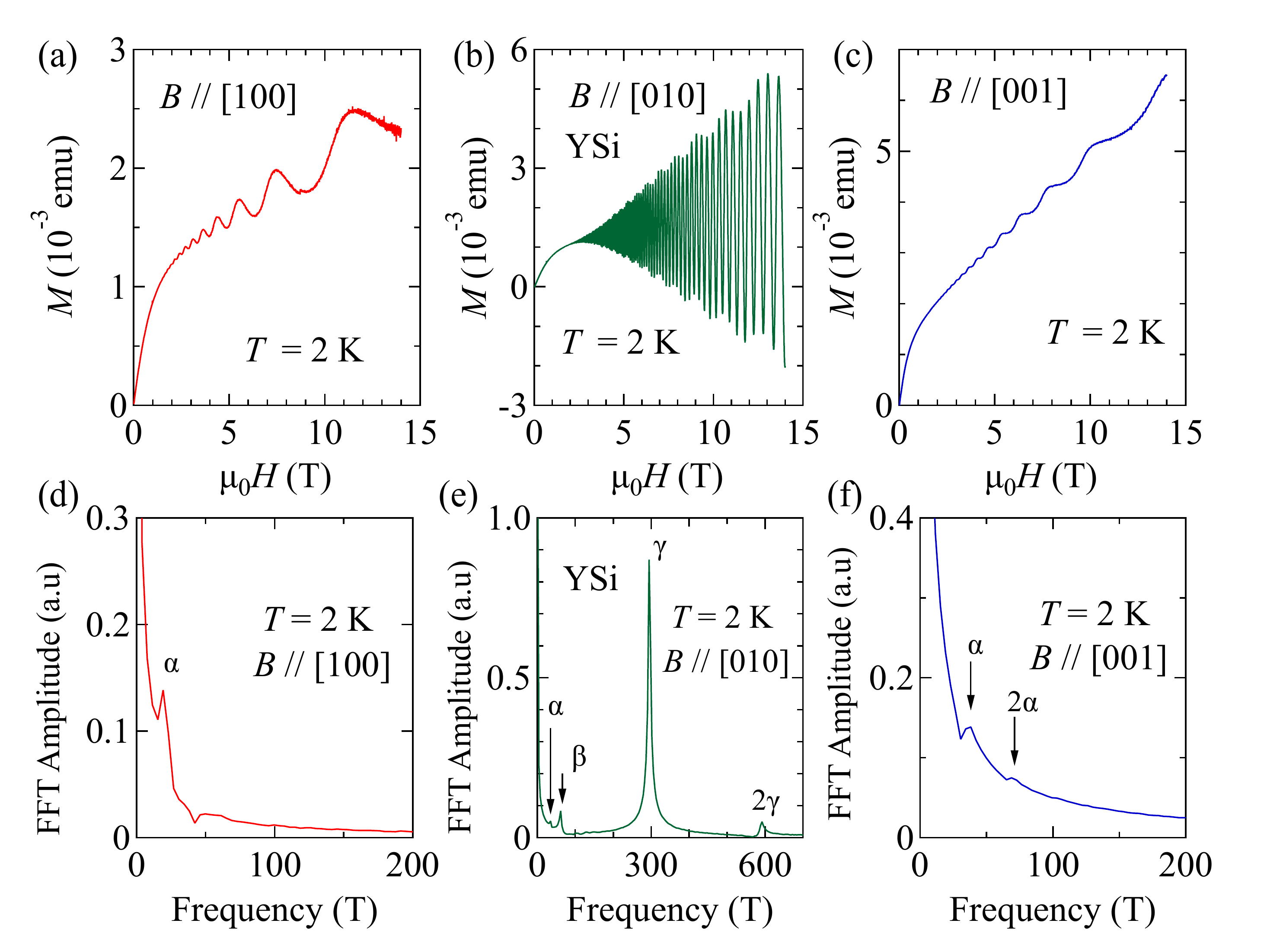}
\caption{(a), (b), and (c) dHvA quantum oscillations observed along the three principal crystallographic directions in YSi at $T =  2K$. (d), (e), and (f) The obtained FFT frequency spectrum. The $\alpha$ branch is observed along all the three directions while $\beta$ and $\gamma$ are observed only along the [010] direction.} 
\label{Fig3}
\end{figure}

The field dependence of the magnetization measurement measured in a VSM at 2~K, up to a field of 14~T, along the three principal crystallographic directions is shown in Fig.~\ref{Fig3}(a), (b) and (c). Robust dHvA oscillations are observed along all three directions. Long-period oscillations are observed for $B~\parallel$~[100] and [001] directions while for [010] direction strong oscillations with multiple frequencies are observed. At $2$~K the oscillations begin to appear from $2$~T field and this corresponds to the magnetic length $l_{\rm B} = \sqrt{\frac{\hbar}{eB}}$~$\approx$~18~nm thus indicating a good quality of the grown single crystal. As the temperature is increased, the amplitudes of the dHvA oscillations tend to decrease and are not discernible for a temperature greater than 21~K. The non-oscillatory background data was subtracted to extract the dHvA oscillation frequency by fast Fourier transform (FFT). The FFT spectra at $T =2$~K along the three principal crystallographic directions are shown in Fig.~\ref{Fig3}(d),(e) and (f). When the magnetic field is along the [100] direction, a single natural frequency at $21$~T has been observed which is named as $\alpha$. For $B~\parallel$~[010] three natural frequencies have been observed at $36$~T ($\alpha$), $61$~T ($\beta$) and $295$~T ($\gamma$) with the second harmonics at $590$~T ($2\gamma$), while for $B~\parallel$~[001] a single natural frequency at $34$~T ($\alpha$) and its corresponding second harmonics was observed at $68$~T ($2\alpha$). It is to be mentioned here that the $\alpha$-branch is observed along all the three directions with a subtle anisotropy in the frequency thus representing a small anisotropic $3D$ Fermi surface. 

\begin{table*}[!]
\centering
  \setlength{\tabcolsep}{8pt} 
	\renewcommand{\arraystretch}{1.5} 
	\caption{Quantum parameters estimated from dHvA oscillations }
	\begin{tabular}{|c |c |c |c |c |c |c |c |c |c |c|} 
		\hline
		B & $F_{exp}$	& $F_{cal}$ & $m^{*}_{exp}$ & $m^{*}_{cal}$ & $\tau_{q}$ 	& $\mu_{q}$ 	& $k_{\mathrm F} $	& $v_{\mathrm F}$	& $l_{q}$ 	& $n$  \\ [0.1ex] 
		& (T)    & (T)     &  $(m_e)$   & $(m_e)$           & $(10^{-13}sec)$  &   $(cm^2V^{-1}s^{-1})$   & (\AA$^{-1})$   & $(10^{5}$m/s)     &   (nm)      &  \\ 
		\hline
		[100] & 21 ($\alpha$) & 21.8 & 0.069(1) & 0.082 &   &       &   0.025   &   4.26    &      &   5.56 *$10^{17}cm^{-3}$\\
		\hline
		[010] & 295 ($\gamma$) & 294.7 & 0.162(2) & 0.147 & 2.2& 2384& 0.095 & 6.75 & 148.65& 7.14 * $10^{12}cm^{-2}$\\

		& 61 ($\beta$) & 71.5 &  0.097(2) & 0.162 & 1.14 & 2053 & 0.043 & 5.1 & 58.39 &1.48*$10^{12}cm^{-2}$ \\
		\hline
		[001] &34 ($\alpha$) & 28.1 & 0.096(4)  & 0.095  & 2.15 & 3927 & 0.032 & 3.85 &82.77 &1.109*$10^{18}cm^{-3}$ \\ 
		\hline

	\end{tabular}
	\label{Tab1}
\end{table*}

\begin{figure}[!]
\centering
\includegraphics[width=0.5\textwidth]{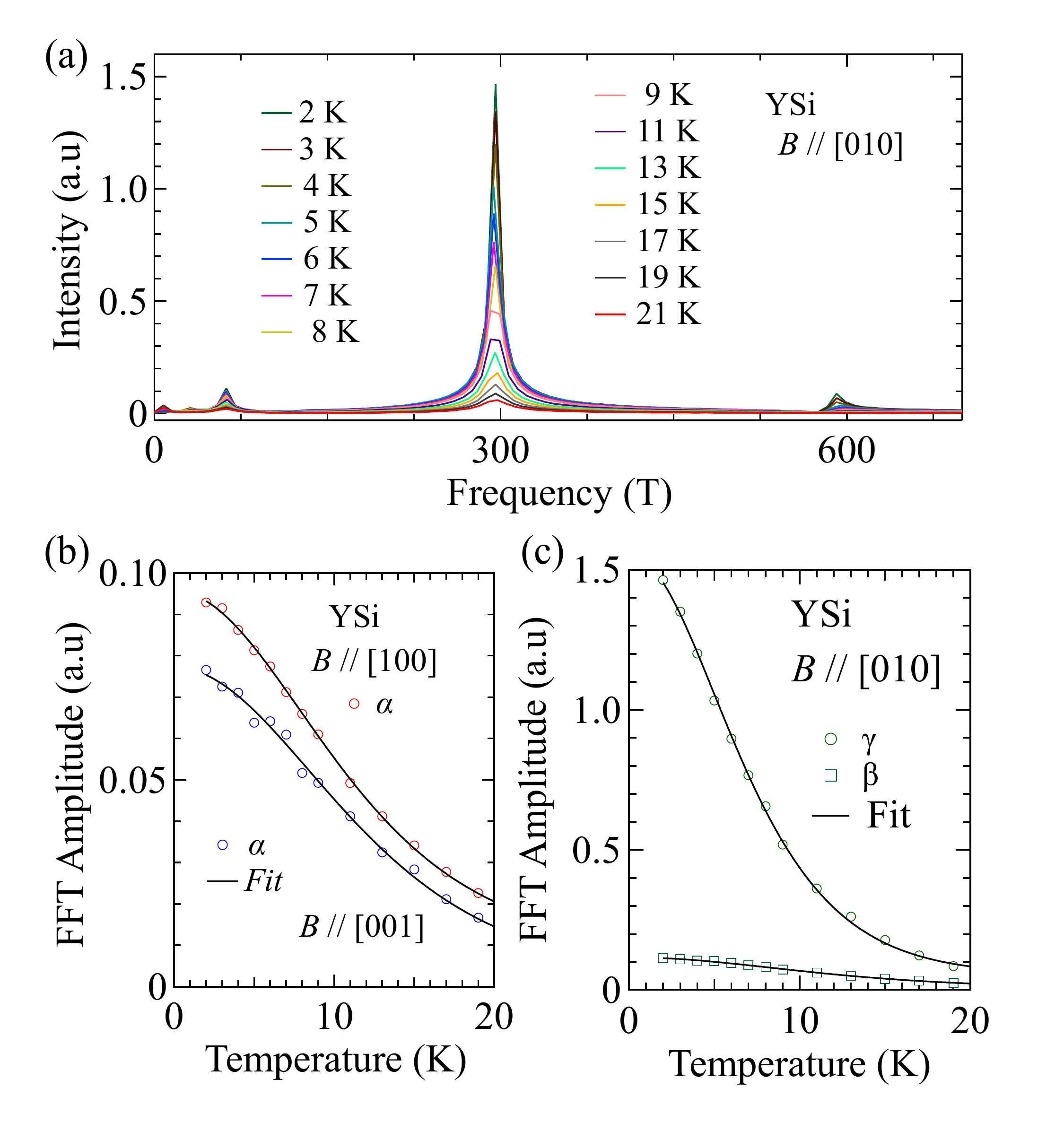}
\caption{(a) Temperature dependence of the FFT amplitude of the oscillation.  The FFT amplitude decreases with increasing temperature.  (b), (c) Mass plot of the frequencies mentioned along the three principal crystallographic directions.  The solid lines are the fits to the thermal damping factor of the Lifshitz-Kosevich expression (see text for details).}
\label{Fig4}
\end{figure}

We have estimated the quantum parameters by analyzing the frequency spectrum of the observed oscillations.  The temperature dependence of the FFT amplitude of the three natural frequencies for $B~\parallel$~[010] is shown in Fig.~\ref{Fig4}(a).  The amplitude of the frequencies decreases as the temperature increases.  The $T$-dependence of the  amplitude of the oscillation frequencies are shown Fig.~\ref{Fig4}(a) and (b) and are fitted to thermal damping factor of Lifshitz-Kosevich expression: $R_{T} = (X/sinhX)$, where $X = (\lambda T m^*/H)$, $\lambda = (2 \pi^2 k_B m_e/e\hbar) (= 14.69$~T) and $m^*$ is the cyclotron effective mass of the charge carriers which is expressed in units of free electron mass $m_e$~\cite{hu2019transport, matin2018extremely, mondal2020extremely} .  The calculated effective masses for the various frequencies are listed in Table~\ref{Tab1}.  It is evident from the table that the effective masses of charge carriers are very small for all the observed main frequencies suggesting Dirac-like dispersion of bands and comparable to that of the gap-less Dirac system Cd$_3$As$_2$~\cite{he2014quantum}.

To understand the topological character of the charge carriers in YSi,  we have performed the Berry phase analysis of the dHvA oscillations using the Lifshitz-Kosevich (L-K) formula~\cite{kumar2017unusual,hu2016evidence} as given below,
\begin{equation}
	\label{Eqn1}
	\Delta M \propto -B^{k}R_{T}R_{D}R_{s}sin\left[2\pi\left(\frac{F}{B}+\psi \right)\right],
\end{equation}  
where $R_T$, $R_D$, and $R_S$ are the thermal, magnetic field and spin damping factors and  $\psi$ is the phase factor.  The expression for $R_T$ has already been defined while that for $R_D$ is $R_D = exp(-14.69 m^* T_{\rm D}/B)$ where $T_{\rm D}$ is the Dingle temperature and $R_{\rm S}$ = $cos(\pi g r m^*/ 2m_0)$, where $\mathtt{g}$ is the Land\'{e} $\mathtt{g}$-factor and $r$ is the harmonic number.   In Eqn.~\ref{Eqn1} $k = 1/2$ for 3D Fermi surface while it is zero for 2D Fermi surface.  The phase factor $\psi$ is given by $\psi$= [$(\frac{1}{2} - \frac{\Phi_B}{2 \pi}) - \delta$], where $\delta$ is the additional phase factor which depends on the dimensionality of the Fermi surface. $\delta = 0$ for $2D$ Fermi surface and $\delta = \pm 1/8$ for $3D$ Fermi surface where the $+$ sign corresponds to  hole pocket and $-$ sign corresponds to  electron pocket.

\begin{figure}[!]
\centering
\includegraphics[width=0.5\textwidth]{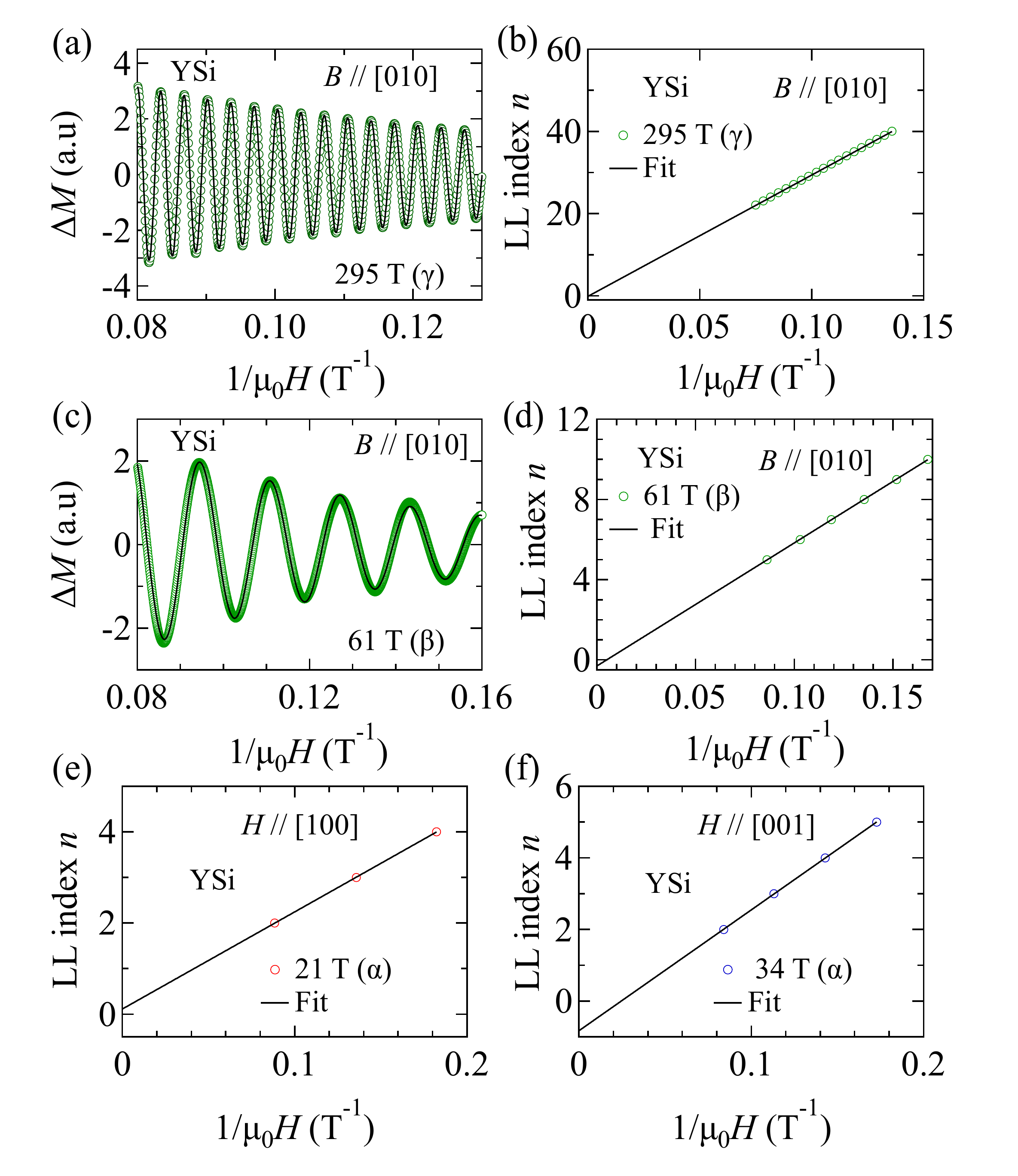}
\caption{(a) The band pass filtered dHvA oscillations of the $\gamma$-branch  measured at $T=2$~K for $B~\parallel$~[010], the solid line corresponds to the L-K formula fitting. (b)  LL-fan diagram corresponding to the $\gamma$-branch. (c) L-K formula fitting of the filtered $\beta$-branch oscillations, (d) the LL-fan diagram corresponding to $\beta$-branch. (e) and (f) LL-fan diagram of the $\alpha$-branch along the [100] and [001] directions.}
	\label{Fig5}
\end{figure}

For  $B~\parallel$~[010] two additional frequencies are observed at 61~T and 295~T  while these two frequencies are absent in the other two directions.  This indicates the anisotropic nature of the Fermi surface and these two frequencies represents a $2D$ Fermi surface.   Using the Onsager's relation $F = \frac{\hbar}{2 \pi e} A_{\rm F}$, we have estimated the cross-sectional area for $\gamma$  frequency as 0.028~\AA$^{-2}$.  In order to estimate the Dingle temperature $T_{\rm D}$  of the $\gamma$-branch, we have used the band-pass filter to isolate the oscillations corresponding to the frequency 295~T and used the $2D$ L-K expression in the high magnetic field region as shown in Fig.~\ref{Fig5}(a).  The reasonably good fit resulted in a Dingle temperature $T_{\rm D}$ as 5.52~K.  A similar estimate of the Dingle temperature was made for the $\beta$  which resulted in $T_{\rm D}$ = 10.62~K.  From the obtained values of $m^*$, $T_{\rm D}$, we have estimated the Fermi wave vector $k_{\rm F}$, Fermi velocity $v_{\rm F}$, quantum scattering $\tau_{\rm q}$, quantum mobility $\mu_{\rm q}$ and the surface carrier density $n_{\rm 2D}$ as listed in Table~\ref{Tab1}.  The information about the Berry phase in YSi has been extracted by plotting the Landau level (LL) fan diagram. We have assigned the LL-index $n (= F/B + \psi)$ to the  minima of quantum oscillations.  For $B~\parallel$~[010] there are multiple frequencies in the quantum oscillation hence we have used the band pass filtered oscillation to construct the LL-fan diagram which is plotted as $n$ as a function of the inverse magnetic field as shown in Fig.~\ref{Fig5}(b) and (d).  The plots are straight line and the slope corresponds to the oscillation frequency while the intercept gives the phase factor $\psi$ mentioned in Eqn.~\ref{Eqn1}.  For the $\gamma$-branch, the intercept is $-0.12$, this amounts to a Berry phase ($\Phi_{\rm B}$) of $1.24\pi$, which reveals the non-trivial character of $\gamma$ pocket.  Similarly, for the $\beta$-branch (61~T) the intercept was estimated to be $-0.3$, this amounts to $\Phi_{\rm B}$ of $1.6~\pi$ suggesting a trivial nature of this band and for the $\alpha$-branch the intercept was $0.114$ and $-0.82$ for $B~\parallel$~[100] and $B~\parallel$~[001] directions (Fig.~\ref{Fig5}(e) and (f)), respectively with $\Phi_{\rm B}$ = $\pi$ and  $2.89\pi$ suggesting a non-trivial nature of the Fermi surface.  For $\alpha$-branch the second-lowest  Landau level has been achieved in a field of $14$~T that exhibits a small Fermi surface which is smaller than the reported Fermi surfaces for Cd$_3$As$_2$ and ${\rm ZrSiS}$ systems~\cite{he2014quantum,  hu2017nearly}.

\section{Conclusion}

We have grown the single crystal of YSi using the Czochralski method. The band structure shows multiple nodal band crossings near the Fermi level. The Fermi surface is formed by both the electron and hole bands, resulting in multi Fermi pockets with both small and large areas.    Importantly, our calculated quantum oscillation frequencies match well with the observed frequencies.  The dHvA oscillation measurements performed along the three principal crystallographic directions have revealed three Fermi pockets, namely, $\alpha$, $\beta$, and $\gamma$, which are well seen in our first-principles calculations. The $\alpha$ band obeys linear dispersion around the Fermi energy which results in a very small effective mass $0.069~m_e$ for $B~\parallel$~[100]-direction and in fact, it is lighter than other popular Dirac and Weyl semimetals~\cite{du2016large, gao2017extremely, butcher2019fermi, wu2019anomalous, singha2017fermi, inamdar2013quantum, chen2018large}.  The estimated Berry phase for this branch is $\pi$ which refers to a non-trivial topological character for this pocket and is observed along all the three principal directions with subtle anisotropy.  Similarly,  $\gamma$-branch also possess very light mass because of linear band dispersion. The observed non-trivial Fermi pocket corresponding to the $\gamma$-branch shows a $2D$ character with a quantum scattering time $2.2~\times~10^{-13}$~s, which is of the same order as observed in ZrSiS and PtBi$_2$ systems~\cite{gao2017extremely, hu2017nearly}. The large quantum mean free path and high quantum mobility signifies the suppression of backscattering in YSi. These results make YSi is an interesting topological material where nontrivial band crossings can be explored in angle-resolved photoemission spectroscopy (ARPES) and scanning tunneling microscopy (STM).

%

\end{document}